\documentclass[twocolumn,prl,superscriptaddress]{revtex4}
\usepackage[utf8]{inputenc}
\usepackage{graphicx}
\usepackage{layouts}
\usepackage{amsmath}
\usepackage{amsfonts}
\usepackage{hyperref}
\usepackage{booktabs} 

\newcommand{\equalcontrib}{\textsuperscript{\textdagger}}

\begin{document}

\title{Deep learning of spatial densities in inhomogeneous correlated quantum systems}

\begin{abstract}

    Machine learning has made important headway in helping to improve the treatment of quantum many-body systems. A domain of particular relevance are correlated inhomogeneous systems. What has been missing so far is a general, scalable deep-learning approach that would enable the rapid prediction of spatial densities for strongly correlated systems in arbitrary potentials. In this work, we present a straightforward scheme, where we learn to predict densities using convolutional neural networks trained on random potentials. While we demonstrate this approach in 1D and 2D lattice models using data from numerical techniques like Quantum Monte Carlo, it is directly applicable as well to training data obtained from experimental quantum simulators. We train networks that can predict the densities of multiple observables simultaneously and that can predict for a whole class of many-body lattice models, for arbitrary system sizes. We show that our approach can handle well the interplay of interference and interactions and the behaviour of models with phase transitions in inhomogeneous situations, and we also illustrate the ability to solve inverse problems, finding a potential for a desired density. Code used to conduct our analysis can be found on GitHub \cite{quant-sic_dmb_2025}.

\end{abstract}

\author{Alex Blania \equalcontrib}
\address{Max Planck Institute for the Science of Light, Staudtstraße 2, 91058 Erlangen, Germany}
\address{IQIM, California Institute of Technology, 1200 E California Blvd, Pasadena, 91125 California, USA}
\address{Physics Department, Friedrich-Alexander-Universit\"at Erlangen-N\"urnberg, Staudtstraße 5, 91058 Erlangen, Germany}
\author{Sandro Herbig \equalcontrib}
\address{Max Planck Institute for the Science of Light, Staudtstraße 2, 91058 Erlangen, Germany}
\address{IQIM, California Institute of Technology, 1200 E California Blvd, Pasadena, 91125 California, USA}
\address{Physics Department, Friedrich-Alexander-Universit\"at Erlangen-N\"urnberg, Staudtstraße 5, 91058 Erlangen, Germany}
\author{Fabian Dechent \equalcontrib}
\address{Humboldt Universität zu Berlin, Institut für Physik, Newtonstraße 15, 12489 Berlin}
\address{Max Planck Institute for the Science of Light, Staudtstraße 2, 91058 Erlangen, Germany}
\author{Evert van Nieuwenburg}
\address{IQIM, California Institute of Technology, 1200 E California Blvd, Pasadena, 91125 California, USA}
\address{Niels Bohr Institute, Blegdamsvej 17, 2100 Copenhagen, Denmark}
\address{Lorentz Institute and Leiden Institute of Advanced Computer Science,
    Leiden University, P.O. Box 9506, 2300 RA Leiden, The Netherlands}
\author{Florian Marquardt}
\address{Max Planck Institute for the Science of Light, Staudtstraße 2, 91058 Erlangen, Germany}
\address{Physics Department, Friedrich-Alexander-Universit\"at Erlangen-N\"urnberg, Staudtstraße 5, 91058 Erlangen, Germany}

\maketitle

\begingroup
\renewcommand\thefootnote{\textdagger}
\footnotetext{These authors contributed equally.}
\addtocounter{footnote}{-1}
\endgroup


Predicting the properties of quantum many-body systems is one of the most important challenges in physics, as it is relevant for quantum chemistry and materials science, as well as for more recent tasks like benchmarking quantum simulators. This is a hard problem for larger system sizes, due to the exponential explosion of the Hilbert space dimension. Elaborate approximation schemes have been developed, including various forms of perturbation theory, mean-field approaches like density-functional theory (DFT), cluster methods, Monte Carlo methods, and several variational-ansatz ideas, including tensor network constructions.
Machine learning has entered this field of predicting quantum matter some years ago, with results that show great promise, turning it into an important tool ~\cite{carrasquillaMachineLearningQuantum2020, carleoMachineLearningPhysical2019, dawidModernApplicationsMachine2022}. It has led to improvements in several directions, like the use of neural networks to represent variational wave functions \cite{carleoSolvingQuantumManybody2017,schmittQuantumManyBodyDynamics2020,hartmannNeuralNetworkApproachDissipative2019}, to improve DFT calculations ~\cite{brockherdeBypassingKohnShamEquations2017, costaDeepLearningDensity2022, chandrasekaranSolvingElectronicStructure2019,ryczkoDeepLearningDensity2019, nelsonMachineLearningDensity2019, morenoDeepLearningHohenbergKohn2019}, as well as in molecular dynamics~\cite{behlerGeneralizedNeuralNetworkRepresentation2007, bartokMachineLearningUnifies2017}.

However, the range of validity of such schemes is often limited, making it necessary to carefully select the approach depending on circumstances. Furthermore, the effort expended in applying the techniques to larger-scale inhomogeneous systems can still be quite substantial. It would, therefore, be desirable to employ machine learning to circumvent entirely these tailored approaches and arrive at a more general method for rapidly predicting the properties of quantum many-body systems. This is especially true in inhomogeneous situations where the need for a scalable approach that can treat large sizes in extended systems arises.

\begin{figure}[htbp]
    \centering
    \includegraphics[width=0.48\textwidth]{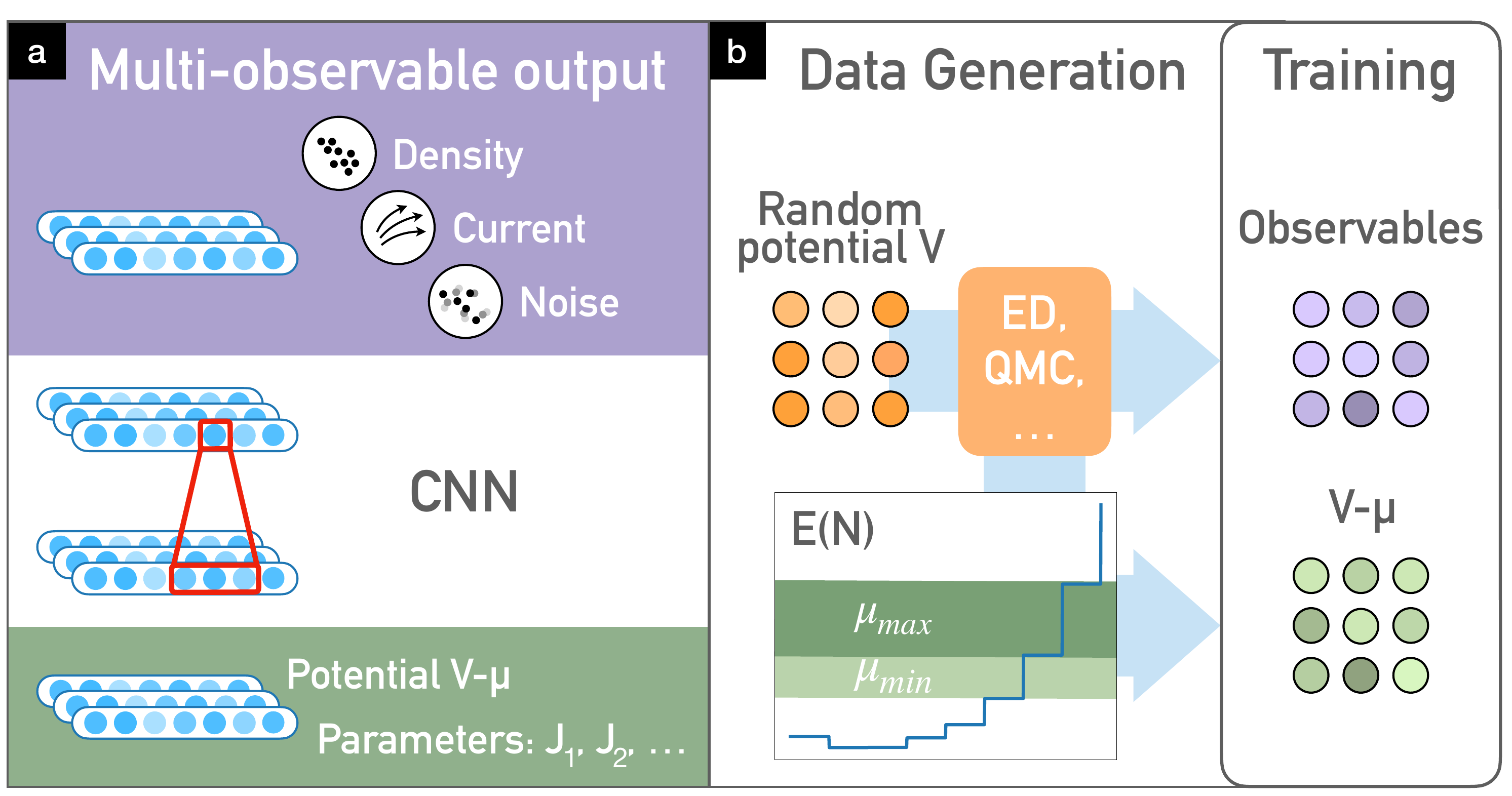}
    \caption{Overview of the approach. (a) The convolutional neural network maps a potential landscape of arbitrary size, together with model parameters, to the spatial maps of multiple observables (including correlators, here denoted as 'noise'). (b) Training procedure. Any of a set of numerical methods are used to turn random potentials into spatial maps, providing the training data. The dependence of energy on particle number is recorded as well and gives access to the chemical potential.}
    \label{fig:overview}
\end{figure}

Here we introduce a straightforward scheme to train a deep convolutional network that is able to directly predict spatial properties of many-body ground states or thermal states in arbitrary potential landscapes.
The key idea is to train on random, spatially non-uniform potentials and for variable particle numbers, where the corresponding predictions can be extracted from any numerical method or potentially even from experimental data.
In general, the network is designed with a multi-headed structure and can thus be trained to produce the spatial maps of several physical observables simultaneously. This is intended to make it more sample-efficient in training, sharing one initial processing block among observables (see Fig \ref{fig:overview}a for a schematic).

In addition, we introduce the capability of training a single network on a range of model parameters, such as coupling values and interaction strengths. This again is designed to boost sample efficiency, since one network is trained on data pertaining to the full parameter range, giving it the potential to make predictions that are partially informed by behaviour seen at other parameter values.

Even though the generation of training data may be computationally expensive and may require a large-scale computing cluster, once the network has been trained our method allows rapid exploration of inhomogeneous quantum many-body systems with low computational effort on single processors; examples show that for typical parameters speedups of a factor of a million are realistic.
Due to the local nature of the convolution, the network can be exploited to make reliable predictions for inhomogeneous systems far larger than those it has been trained on previously, provided that training sample sizes and the combined range of the stacked convolutions exceed the physical correlation length.
One potential application of the trained network would be Hamiltonian learning~\cite{valentiHamiltonianLearningQuantum2019a}, finding underlying parameters and potentials given experimentally observed densities via gradient descent. Another application would be the inverse design of potentials with desired characteristics, either via exploiting the rapid predictions in gradient-free methods or implementing direct gradient descent.

In previous works, scalable approaches based on convolutional or recurrent neural networks have been proposed in the context of quantum many-body systems. Those approaches map a potential to a scalar global variable like the energy \cite{saraceniScalableNeuralNetworks2020,mujalSupervisedLearningFew2021} or to level-spacing statistics \cite{greschScalableApproachManybody2022}. In contrast, our approach maps a potential to several spatial distributions of observables like the density. Although there have also been some works that investigate the direct prediction of densities, those were in the context of small few-particle quantum-chemistry type scenarios \cite{ryczkoDeepLearningDensity2019}, where scalability and application to an extensive system was not an issue.

We consider interacting lattice models of fermions or bosons, subject to a potential landscape $V(x)$, where $x$ denotes the discrete lattice location. Our aim is to train a deep neural network to be able to predict the spatial dependence of the ground-state expectation values of interesting observables when it is provided with the potential.
These observables may include the density ${\hat n}(x)={\hat a}^{\dagger}(x) {\hat a}(x)$, the coherences that enter the density matrix ${\hat a}^{\dagger} (x+\delta x) {\hat a}(x)$ (which can be used to express the current flow if $x+\delta x$ denotes a neighboring site), or fluctuations like ${\hat n}(x+\delta x) {\hat n}(x)$.

In our approach, the starting point for training is the evaluation of such spatial observable maps by any of a number of numerical techniques, like exact diagonalization (ED) or quantum Monte Carlo (QMC).
Training samples are produced by generating random potentials and calculating predictions for those.
We take these potentials to be realizations of both white and colored noise, with similar results.
Samples are produced on relatively small finite-size systems of linear extent $L$, and for different choices of particle number $N$.
This idea is related to a recent approach trying to learn the dynamics of quantum many-body systems by observing their behaviour under random {\em temporal driving} \cite{mohseniDeepLearningQuantum2022}.

One important characteristic of our approach is that we are interested in using the network to produce predictions for \emph{arbitrary-size systems} later on. In order to handle that, we choose to work in the grand-canonical ensemble, where the particle number can be tuned via the chemical potential $\mu$.
Instead of taking $V(x), N, L$ as input, we hence feed $V(x)-\mu$ into the network.
To facilitate this, during the sample generation phase we scan the evolution of ground-state energy $E$ vs. particle number $N$ and extract $\mu$ from the resulting dependency. Specifically, for each sample fed into the network, given $E(N)$ we pick \(\mu\) drawn uniformly from the interval \(\left[E(N) - E(N-1), E(N+1)-E(N) \right]\).
This ansatz effectively produces an interpolated version of $\mu(N)$ and leads to the desired ability to scale to arbitrary sizes $L$. For the specific QMC method we use in our studies, $\mu$ can be prescribed directly (see Fig.~\ref{fig:overview}b for a summary of the described procedure).

One central aim in selecting the architecture of our network is to make sure to maximize the sample efficiency, i.e. to exploit as much as possible the information contained in the training data, which is expensive to generate. In general, the architecture of the convolutional network is set up with a multi-channel output structure that is asked to predict multiple observable maps simultaneously. This enables more efficient use of the training data (as compared to training several different observable-specific networks), since processing steps inside the network are shared for the different observables, and weight updates benefit from these different sources of training information. The input to the network, as stated above, is the potential $V(x)-\mu$ on a lattice, provided in the shape of an 'image' in $d$ dimensions. Furthermore, we may often be interested in predicting not for a fixed model, but for a whole class of quantum many-body models, specified via a Hamiltonian ${\hat H}(J_1,J_2,\ldots)$ with variable coupling parameters $J_1,J_2,\ldots$. To this end, we feed extra 'channels' to the network, one channel for each parameter. For simplicity these are taken to be of the same spatial extent as the potential, but defined to be constant throughout space (this setup would also allow a straightforward treatment of spatial inhomogeneities in such parameters). Training on randomly chosen parameter combinations within some parameter range again maximizes sample efficiency, i.e. it is more economical than training a number of parameter-specific networks, since behaviour learned in other regions of parameter space can often be extrapolated. As to the specifics of the convolutional neural network, many favourite choices from deep image classification and processing are possible. In the numerical investigations discussed below, we used both plain convolutional networks as well as established architectures like the ResNet \cite{heDeepResidualLearning2015}. We augmented the limited datasets through symmetry transformations (such as rotations and reflections).

One of the defining characteristics of quantum many-body problems in inhomogeneous potentials is the interplay of interference and interactions. Interference is generated when matter waves scatter from features in the potential, but interactions can modify this interference in multiple ways -- for example, they might smear out interference features via fluctuations, change the wave length in a density-dependent manner, or even enhance the density contrast of interference fringes. Situations where both interference and interactions are present thus can serve as a natural benchmark case for assessing the quality of predictions for any new numerical or approximate technique designed to handle generic inhomogeneous many-body systems.

\begin{figure}[htbp]
    \centering
    \includegraphics[width=0.48\textwidth]{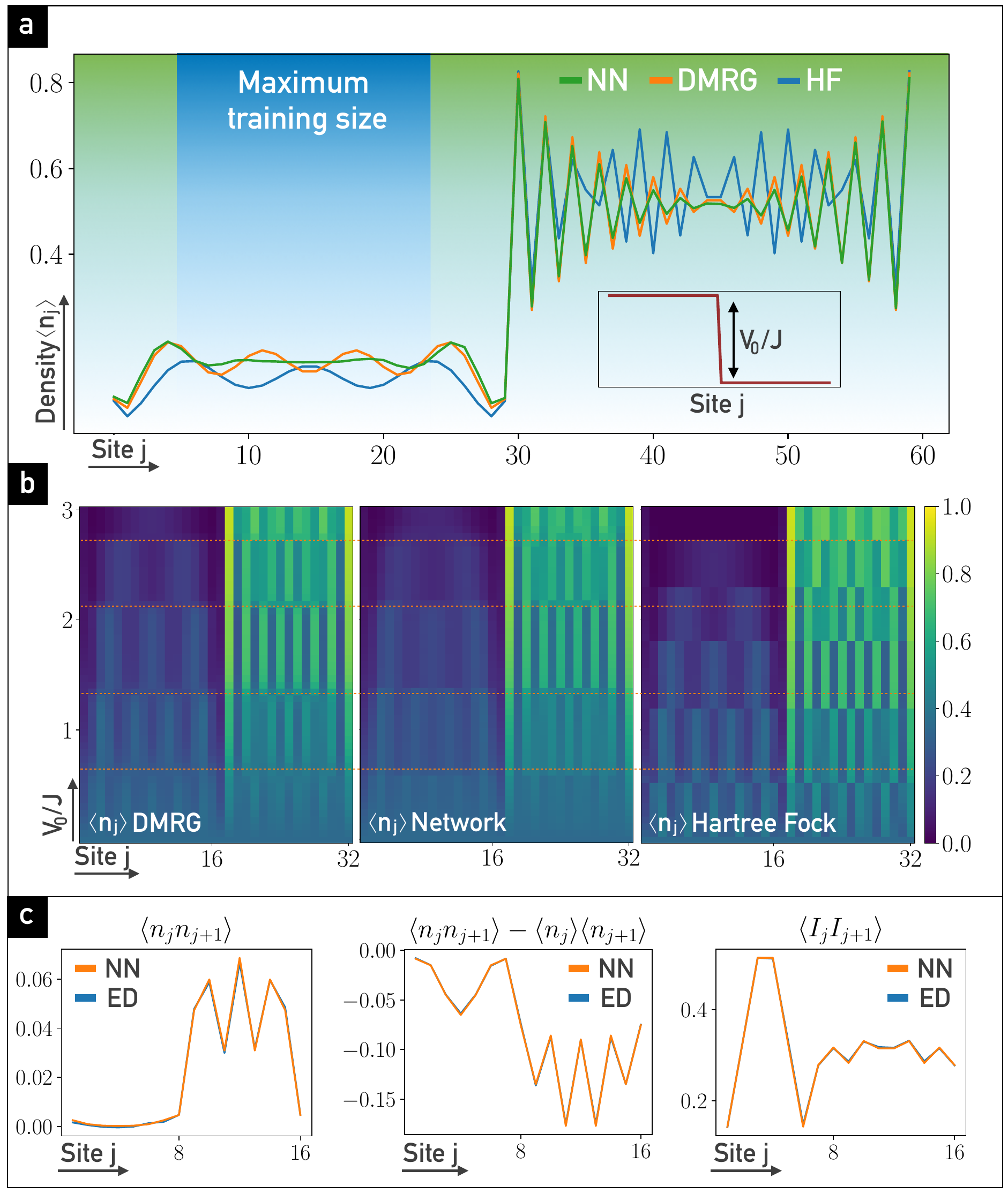}
    \caption{Performance of the network addressing the interplay of interactions and interference. (a) Density for a step potential (never seen during training), displaying Friedel oscillations in a 1D system of interacting fermions. The NN prediction is very close to the almost exact result (DMRG), in contrast to Hartree-Fock. (b) Density evolution vs. potential step height, comparing the three different approaches; again the NN performs very well. (c) Predictions for multiple observables, obtained from the single (multi-head) network, evaluated for a step potential in a smaller system size. We compare the network output with the exact diagonalization (ED) result.}
    \label{fig:friedel_1d}
\end{figure}

We chose to study these effects in an interacting fermionic 1D lattice system. One-dimensional systems are good for benchmark comparisons, since we can obtain essentially exact results on rather large systems. At the same time, and for this very reason, 1D systems are not necessarily the primary target for eventual applications of our network-based approach, with the possible exception of inverse problems. Nevertheless, we find that the performance of the deep learning approach is not very sensitive to the dimensionality, in contrast to other approximate and numerical techniques. Therefore, in our assessment the comparisons obtained in 1D are meaningful to indicate the overall accuracy of the method independent of dimensionality, which is confirmed by our tests in 2D, discussed further below. The specific model we selected describes spinless fermions and contains nearest-neighbor interactions:
\begin{equation}
    \begin{split}
        \hat H = & - J \sum_x {\hat a}^{\dagger}(x) {\hat a}(x+1) + \mathit{h.c.}                    \\
                 & + U \sum_x {\hat n}(x) {\hat n}(x+1) + \sum_x V(x) {\hat n}(x), \label{eq-1D-Ham}
    \end{split}
\end{equation}
where \(U\) denotes the interaction strength and $J$ the hopping amplitude, and we choose periodic boundary conditions. After training on random potentials with extent \(L\in \{5,\ldots,16\}\) (in this case, for fixed parameters), as described above, we ask the network to predict the density for a step-wise potential in the form of a potential well. This kind of potential generates Friedel oscillations with a density-dependent wavelength and an asymptotic power-law decay modified by the interactions, as suggested by Luttinger liquid theory. To quantify the accuracy of the neural-network predictions on this test case, we compare them against results from two other techniques: density-matrix renormalization group (DMRG) (code taken from \cite{fishmanITensorSoftwareLibrary2022}), which is essentially exact, and Hartree-Fock. The results are shown in Fig.\ref{fig:friedel_1d}.

We observe a very good match with DMRG predictions, while Hartree-Fock struggles and gives unreliable results. This is particularly apparent when sweeping the potential step height. That leads to characteristic step-wise changes in the total particle number contained in the potential well and resulting changes in the density-dependent wave length of the oscillations. These are captured well by the network -- even though to do so the network has to extract non-local information, since the particle number quantization depends on the extent of the well.
In this example of interacting 1D fermionic systems, we also take the opportunity to train and predict multiple observables, as recounted above. In addition to observables connected to the density \(\hat{n}\), we predict the current between sites, \(\hat{I}(x) = i J (\hat{a}^\dagger(x-\delta x)\hat{a}(x) - \hat{a}^\dagger(x)\hat{a}(x-\delta x))\), and its correlator. In this case, we compare the NN predictions vs. exact diagonalization (choosing a smaller lattice size for that purpose) and observe an extremely close match (Fig.~\ref{fig:friedel_1d}c). We have verified that training samples are not close to the test potential.

The second main characteristic of inhomogeneous quantum many-body systems, beyond the interplay of interference and interactions, is the appearance of phase transitions and the resulting spatial patterns of different phases produced in an external potential. A correct predictive approach must not only reproduce the homogeneous bulk phase diagram but also be able to capture features in inhomogeneous scenarios, like finite penetration depths observed at domains walls separating different phases in space and the strongly modified response to local perturbations when the system is close to a phase boundary.

We have investigated the performance of our deep-learning approach for a paradigmatic 2D model system, the bosonic Hubbard model. Specifically, we selected the extended model with both onsite and nearest-neighbor interactions. This model~\cite{ohgoeGroundStatePhaseDiagram2012,sutharSupersolidPhaseExtended2020} is known to have multiple phases, including the homogeneous Mott insulator and superfluid, but also phases with a spontaneously arising periodic density modulation (checkerboard and supersolid phases). The Hamiltonian for this model is given by
\begin{equation}\label{eq-2D-Ham}
    \begin{split}
        {\hat H} = & -J \sum_{x,x'} {\hat a}^{\dagger}(x) {\hat a}(x')                           \\
                   & + {U \over 2} \sum_x {\hat n}(x) ({\hat n}(x)-1)                            \\
                   & + U' \sum_{x,x'} {\hat n}(x) {\hat n}(x') +  \sum_x (V(x)-\mu) {\hat n}(x).
    \end{split}
\end{equation}
%
%
Here $U$ and $U'$ are the onsite and nearest-neighbor interactions, respectively, and the sum over $x'$ is assumed to range over all nearest neighbors of $x$. In this example, we want to learn not only for a given model, with fixed interactions, but rather be able to make predictions for a whole class of models. To this end, during training we choose $U$ and $U'$ for each sample randomly, within certain ranges. As explained above, the values of these parameters are fed into the network in the form of spatially constant additional input channels. By training in this way on a range of parameters, the network profits from its ability to interpolate, i.e. all the observed samples will provide useful information even for predictions eventually extracted at other parameter combinations.

A particular challenge for NN training is produced when a phase has a spontaneously broken symmetry, since the predictions for a given potential can assume one of several outcomes, which would confuse the network.
This general aspect is illustrated nicely for the case of the checkerboard solid, which shows a staggered density. In contrast, the spontaneously broken U(1) symmetry for the superfluid does not affect the observables we consider. To cope with this challenge,
during training we additionally supply the network with an extra channel, that effectively selects the broken symmetry sector. Specifically for the checkerboard, we provide another channel as input to the network, which contains the checkerboard variant (one of two), which is most strongly correlated with the observed density. This is even necessary for inhomogeneous potentials, since QMC does not reliably pick the energetically slightly favourable variant.

We train on predictions obtained using finite-temperature QMC on relatively small systems with an even number of sites \(L \in \{8,10,12,14,16,18,20\}\) using the non-uniform worm algorithm~\cite{sadoune_efficient_2022}. For the training, we create data sets of up to 20,000 samples, though we checked that already 2,000 samples lead to very good results (see Fig. \ref{fig:bose_hubbard_2d}f). Further details about the data generation can be found in the appendix. As labels we use 6 observables derived from the density \(\langle n_{ij} \rangle\), \(\langle n_{ij} n_{i,j+1} \rangle\), \(\langle n_{ij} n_{i+1,j} \rangle\), \(\langle n_{ij} n_{i+1,j+1} \rangle\), \(\langle n_{ij} n_{i+1,j-1} \rangle\) and \(\langle n_{ij}^2\rangle\).

As training is performed on random potentials, a non-trivial check on the accuracy of the resulting trained network can already be obtained by using the network to predict the phases of homogeneous situations. The resulting phase diagram is shown in Fig.~\ref{fig:bose_hubbard_2d}a, with a very good match against QMC predictions. Moving beyond bulk phases, we test predictions for inhomogeneous situations. In the first example, we choose a parabolic potential underlying the lattice, which is a typical situation in experiments with ultracold atoms in optical lattices. The resulting characteristic stepwise density distribution (in the case of the standard bosonic Hubbard model with Mott and superfluid phases) is known as a 'wedding cake' structure. In the present case as shown in Fig.~\ref{fig:bose_hubbard_2d}d, there are more phases visible, and we compare against QMC predictions. In a second example depicted in Fig.~\ref{fig:bose_hubbard_2d}c, we take a square potential well, and we study the resulting density distribution vs. a varying potential depth, again performing a comparison against the numerically exact QMC result.

\begin{figure*}[htbp]
    \centering
    \includegraphics[width=0.99\textwidth]{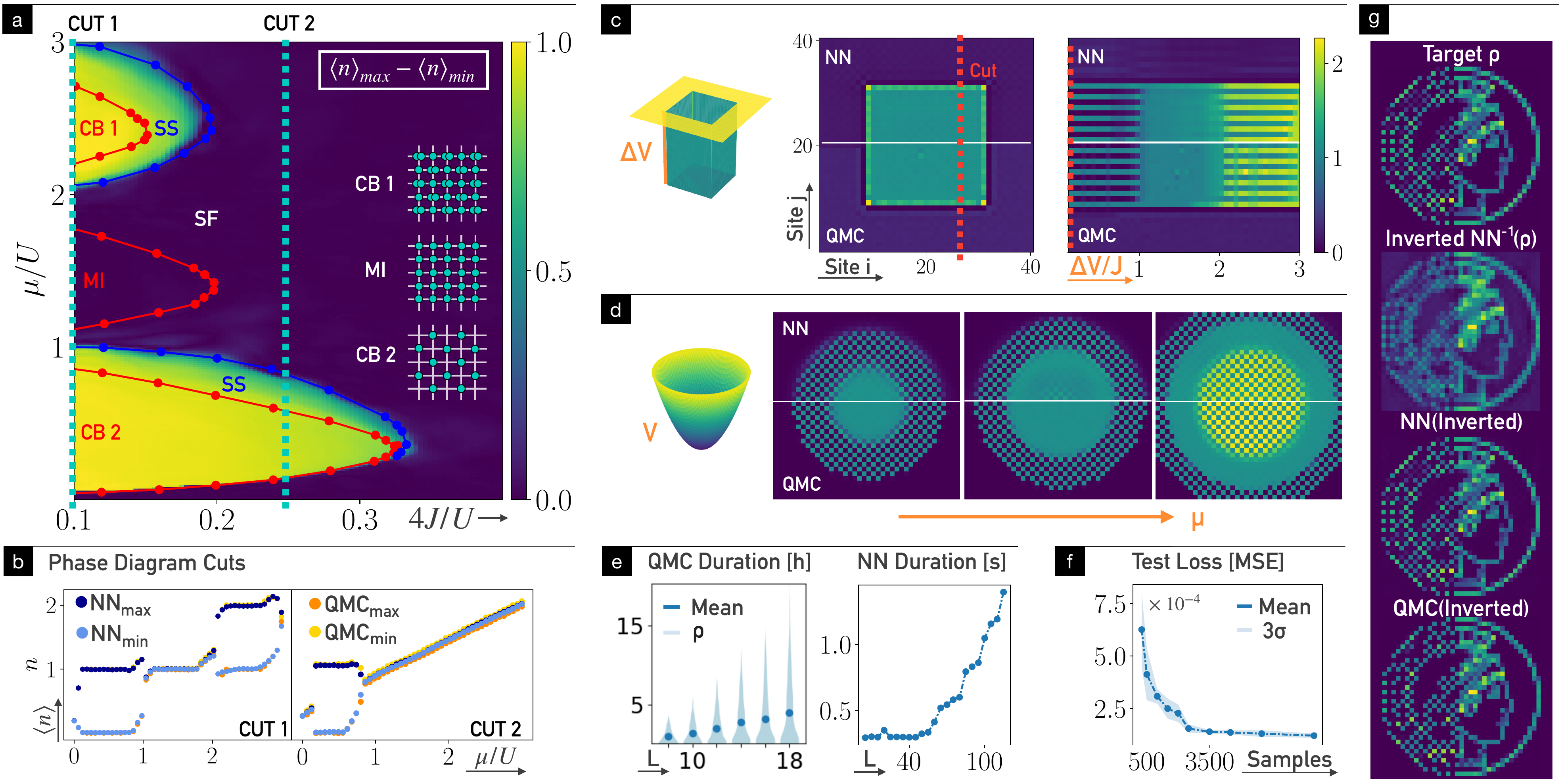}
    \caption{Results for a 2D model with phase transitions. (a) Phase diagram for the 2D Hubbard model with nearest-neighbor coupling, vs. chemical potential and hopping strength for \(4U'/U = 1.0\). The NN (trained on finite-T QMC results) is asked to predict the density for a flat potential, after having been trained on random potentials. We plot the difference between the maximum and the minimum of the density. The indicated phase boundaries mark the ground-state phase transitions extracted from \cite{ohgoeGroundStatePhaseDiagram2012}. Checkerboard solids \(\textrm{CB}_1\) and \(\textrm{CB}_2\) with filling factors \(\rho=3/2,1/2\), supersolid (SS), superfluid (SF), Mott insulator (MI, \(\rho = 1\)). (b) Cuts through (a), comparing predictions of the network with those obtained from QMC (with max. and min. densities). QMC densities in the SF phase are not perfectly constant in space, due to sampling noise. (c) Predictions (NN vs. QMC) for a 2D potential well. Right plot shows evolution of density along a 1D cut for varying well depth. (d) Predictions for a harmonic potential, changing the chemical potential.
        (e) Wall-clock time for QMC calculations vs. system size (for a fixed convergence criterion), and the same for NN evaluation (inference time). (f) Test error (mean squared error) vs. training set size, indicating fluctuations (standard deviation $\sigma$) over several training runs. (g) For given parameters (4J/U=0.25,zU'/U=1.0) the network (NN) is used to "invert" a prescribed density \(\rho\) to produce a corresponding potential (Inverted), see main text, which is tested by using both the NN and QMC for recovering the density.
    }
    \label{fig:bose_hubbard_2d}
\end{figure*}

Having trained a neural network for the task of predicting inhomogeneous quantum many-body systems, the most straightforward benefit is speed, enabling application to large-scale systems. We illustrate this by comparing the time it takes to run QMC simulations vs. the time it takes the neural network to produce predictions, as a function of the system size (Fig.~\ref{fig:bose_hubbard_2d}e). Even though such a comparison comes with caveats -- e.g. the time needed to run QMC for a given potential depends on the details of the convergence criteria adopted -- it is clear that there is a difference of many orders of magnitude, and this observation is independent of such details. Specifically for system size 40, we achieve a speed up of a factor of \(5\cdot 10^{5}\), which is likely even more pronounced for larger systems, as the NN inference time scales much more favorably with the system size. Another important aspect, besides inference time, is the time it takes to generate training samples using expensive techniques like QMC. In Fig.~\ref{fig:bose_hubbard_2d}f, we show the test loss as a function of the number of training samples, observing that even for a few thousand samples the loss is already on the same order of magnitude as for $10^5$ samples. This is a very important message for the feasibility of the whole method.

Another important benefit of using a network is that it represents a differentiable mapping from input to output. This can be exploited for inverse design. For that purpose, we would start from a given desired observable map (e.g. a density distribution), or, more generally, some cost function that prefers certain features in the predicted observable map. We can then take the derivative of that cost function with respect to the input, i.e. with respect to the underlying potential (or the model parameters), in order to do gradient descent and find the potential landscape that is able to produce the desired effect. Here we constrain the inversion to ranges the network was previously trained on, by applying gradient descent to an unconstrained input variable, which is mapped to a constrained network input by an accordingly scaled and shifted sigmoid function. In some cases, there may be a continuum of possible solutions, in other cases - possibly due to the constrained nature of the inversion - one might only find a compromise solution, i.e. come as close as possible to the desired output. We illustrate this use case in Fig.~\ref{fig:bose_hubbard_2d}g.

In conclusion, we have shown how a deep neural network can efficiently predict inhomogeneous densities in large-scale quantum many-body systems, including aspects like simultaneous treatment of multiple observables. With this technique, one can speed up the exploration of designed potentials by several orders of magnitude and apply inverse design to obtain desired outcomes. Another promising future application would be Hamiltonian learning, deducing the potential and model parameters from experimental observations in the ground state of a system. While we have trained on simulations, modern developments will enable training on experimental data, most notably using quantum gas microscopes for optical lattices which also enable local potential programming \cite{gross2021quantum}.

\bibliographystyle{apsrev4-2}
\bibliography{bib/NNManyBodyDensity,bib/FinalManuscriptExtraBib}

\section{Appendix}

\subsection{Training Data}\label{sec:data_gen}

In the 1D examples, we trained on potentials $V(x)$ that are realizations of uniform white noise (\(\overline{V(x)V(x')} = \delta_{x,x'}\)) with \(V(x)/J \in[-12,12]\). The input for the network was solely the combination \(V(x)-\mu\), where \(\mu\) is extracted from the energy dependence on the number of particles, as described in the main text.

For the 2D example, we instead used Gaussian Random Fields with periodic boundary conditions with a randomly selected power spectrum. For all allowed discrete frequencies in Fourier space at point \(k_i,k_j\) a complex amplitude \(a_{ij}\) is drawn. For a randomly drawn exponent \(n\) from a log-uniform distribution from the interval [0.1,10], the amplitude is then multiplied by \( {\left( \sqrt{k_i^2 + k_j^2} \right)}^{-n} \). Finally the real part of the inverse Fourier transform of the array of frequencies is randomly rescaled such that the absolute maximum of V, \( 0.3 \leq \max(|V|)/|\mu| \leq 2 \).
In addition, in the 2D example, we supply the model parameters in the Hamiltonian as input channels to the network.
For the training set, these parameters were drawn from uniform distributions. In particular, we selected \(4J/U\in [0.05,1]\), where the hopping strength \(J\) is set to one, \(4U'/U \in [0.75,1.75]\) and \(\mu \in [-0.5U,3U]\).

To simulate the density for the Bose Hubbard model for bosons in 2D, we adapt the code from  \cite{worm_pollet_code,worm_pollet} to include the possibility for non-uniform input potentials, as well as store all measurements necessary to compute the output observables. All simulations are run with the parameters \(C_{worm}=2.0\), \(p_{insertworm}=1.0\), \(p_{moveworm}=0.3\), \(p_{insertkink}=0.2\), \(p_{deletekink}=0.2\), \(p_{glueworm}=0.3\), \(\beta=20 \) and \(n_{max}=3\).

The integrated autocorrelation \( \tau_{int} \) varies drastically between potential regimes. This requires a robust tuning procedure of measurement intervals. In an iterative manner, the number of samples between measurements is increased exponentially \(\{35\cdot 1.8^i | i \in [0,16] \}\), until \(\tau_{int}\), calculated using \cite{WOLFF2004143,de_palma_python_2019}, is estimated to be smaller than 25. With the resulting measurement interval, the simulation is run, until the absolute error on the density \(\langle \hat{n}_{ij}\rangle \), is smaller than 0.015. For efficiency intermediate tuning runs are skipped, if the evaluated \( \tau_{int}\) is much larger than desired.

\subsection{Network Architecture}

For both experiments convolutional, residual networks \cite{heDeepResidualLearning2016b} are used. Their application is inherently local, which enables the prediction of densities independent of potential lattice size. To mimic periodic boundary conditions, the convolutions use a periodic padding.
The resolution of the input is left unchanged throughout the network. Since the training was done with small potential lattices, alternatives like U-nets \cite{10.1007/978-3-319-24574-4_28} were not considered and no downsampling in the spatial dimensions is applied. To map to the predicted observables a single convolution is used.

For the 1D model, we used a correspondingly modified ResNet-type architecture with 64 blocks. Every layer produces 120 output channels.

\begin{figure}[hbtp]
    \centering
    \includegraphics[width=\linewidth]{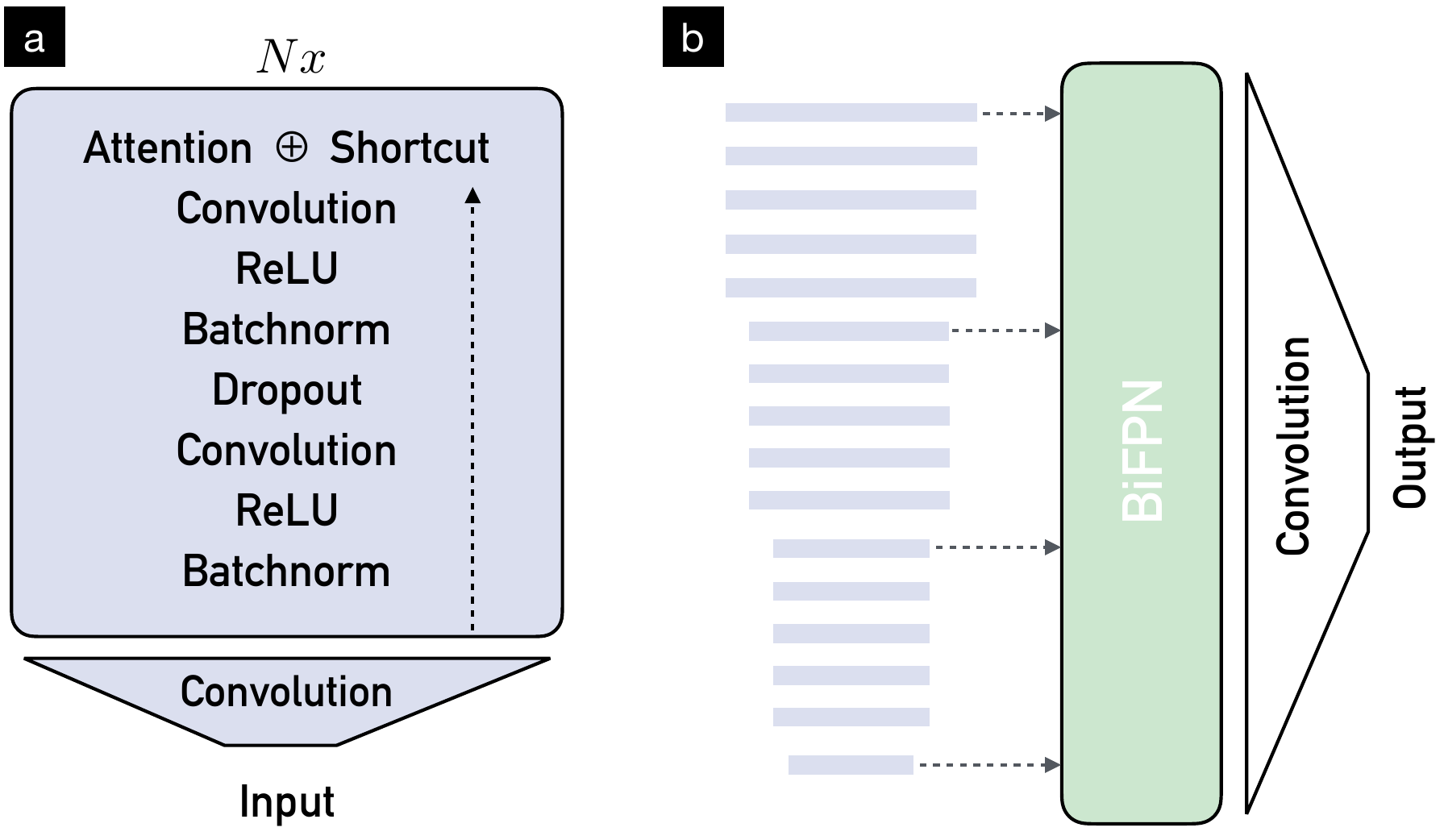}
    \caption{Architecture of the 2D model. a) The residual blocks take a preactivation structure \cite{10.1007/978-3-319-46493-0_38}. b) Overall network architecture with residual blocks (purple) and BiFPN structure.}
    \label{fig:2d_network_architecture}
\end{figure}

For the 2D experiment additional network components were added. We use a backbone and BiFPN \cite{9156454} weighted feature fusion blocks. The backbone architecture as well as the complete network are depicted in Fig.\ref{fig:2d_network_architecture}. For attention, we use CBAM \cite{Woo_2018_ECCV}, with the modification that the spatial component attends to average, max and standard deviation poolings.

The shortcut is either the identity, if the \#input channels is equal to \#output channels. Otherwise a convolution with kernel size 1, followed by a Batchnorm \cite{ioffeBatchNormalizationAccelerating2015c}.

The probability for dropout is set to 0.1. We use ReLU for non-linearities. Layer dimensions are listed in Table\ref{tab:2d_network_dimensions}.

\begin{table}[htbp]
    \caption{2D network dimensions.}
    \label{tab:2d_network_dimensions}
    \begin{ruledtabular} 
        \begin{tabular}{cccc}
            \textbf{Channels} & \textbf{Blocks}         & \textbf{Kernel} & \textbf{BiFPN}           \\
            \hline 
            32                & 1 block (index 0)       & 3$\times$3      & \checkmark               \\
            64                & 3 blocks (indices 1-3)  & 3$\times$3      & \checkmark (at index 3)  \\
            128               & 5 blocks (indices 4-8)  & 3$\times$3      & \checkmark (at index 8)  \\
            256               & 7 blocks (indices 9-15) & 3$\times$3      & \checkmark (at index 15) \\
        \end{tabular}
    \end{ruledtabular}
\end{table}

In order to include the positivity of the predicted observables manifestly into the model, an exponential function is applied to the neural network output.

\subsection{Training Setup}

For training, we use common techniques such as gradient clipping (at L2-norm 1.0), stochastic weight averaging \cite{izmailov2018averaging} and gradient accumulation. We use the AdamW optimizer \cite{LoshchilovH19} with a base learning rate of \( 5\cdot 10^{-5}\) with ReduceLROnPlateau learning rate scheduling.

For the 2D experiment, we use a batch size of 512 and accumulate 8 batches for each update step.

During training, each batch is drawn uniformly from all samples irrespective of sample lattice size. The model maps each mini batch of samples onto the predicted observables and the losses for all lattice sizes are aggregated as a sum.

The model application should be equivariant with respect to rotations and flips of the input potential in the 2D case and flips in the 1D case. However, in contrast to other work (eg. \cite{cesa2022a}) our network architecture is not manifestly equivariant. We thus randomly rotate and flip the input and label during training. We also add a loss term, which optimizes for equivariance. Given an input \( \bf{x} \in \mathbb{R}^{ F \times
    H\times W} \), \( g \in \mathrm{D}_4 \),
\begin{equation*}
    \mathcal{L}_{E}(\bf{x})=\mathrm{Var}\left[ \left\{ \phi_g^{-1}\left(f_{\bf{\Theta}} \left( \phi_g( \bf{x} ) \right) \right) | g \in \mathrm{D}_4 \right\}  \right],
\end{equation*}

where \( \phi_g \) is a mapping of the indices of \(\bf{x}\) based on the group element \( g\). During training we estimate the variance based on 8 (2D) or 2 (1D) randomly mapped augmentations for each unique input sample in the batch.

For the main loss term, we use the mean absolute error (MAE) for the 1D and mean squared logarithmic error (MSLE) for the 2D experiment, as all observables are positive.

To accelerate backpropagation during training with MSLE, we note that since \(\bf{y}_{pred}=\exp(\bf{y}'_{pred})\), ie. an exponential applied to the raw network output, we have \(\ln(1+\bf{y}_{pred})= \ln(1 + \exp(\mathbf{y}'_{pred}))\). This is the softplus function applied to the raw network output. For loss computations, we thus apply the softplus:
\begin{equation*}
    \bf{y}_{pred, loss} = \ln( \mathbf{1} + \exp(\bf{y}_{pred}'))
\end{equation*}
to the model output \(\bf{y}_{pred}'\) and \(\operatorname{\ln1p}\) to the label:
\begin{equation*}
    \bf{y}_{label,loss} = \operatorname{\ln1p}(\bf{y}_{label}) = \ln(1 + \bf{y}_{label}).
\end{equation*}
Both values are then fed into an MSE loss, which is mathematically equivalent to an MSLE loss:
\begin{align*}
    \mathcal{L}_{MSLE}(\mathbf{y}_{\text{pred}}, \mathbf{y}_{\text{label}})
     & = \mathbb{E} \Big[
    \big( \ln(1 + \mathbf{y}_{\text{pred}})               \\
     & \quad - \ln(1 + \mathbf{y}_{\text{label}}) \big)^2
    \Big],                                                \\
    \mathcal{L}_{MSE}(\mathbf{y}_{\text{pred,loss}}, \mathbf{y}_{\text{label,loss}})
     & = \mathbb{E} \Big[
        \big( \mathbf{y}_{\text{pred,loss}} - \mathbf{y}_{\text{label,loss}} \big)^2
        \Big].
\end{align*}

\begin{figure}[hbtp]
    \centering
    \includegraphics[width=\linewidth]{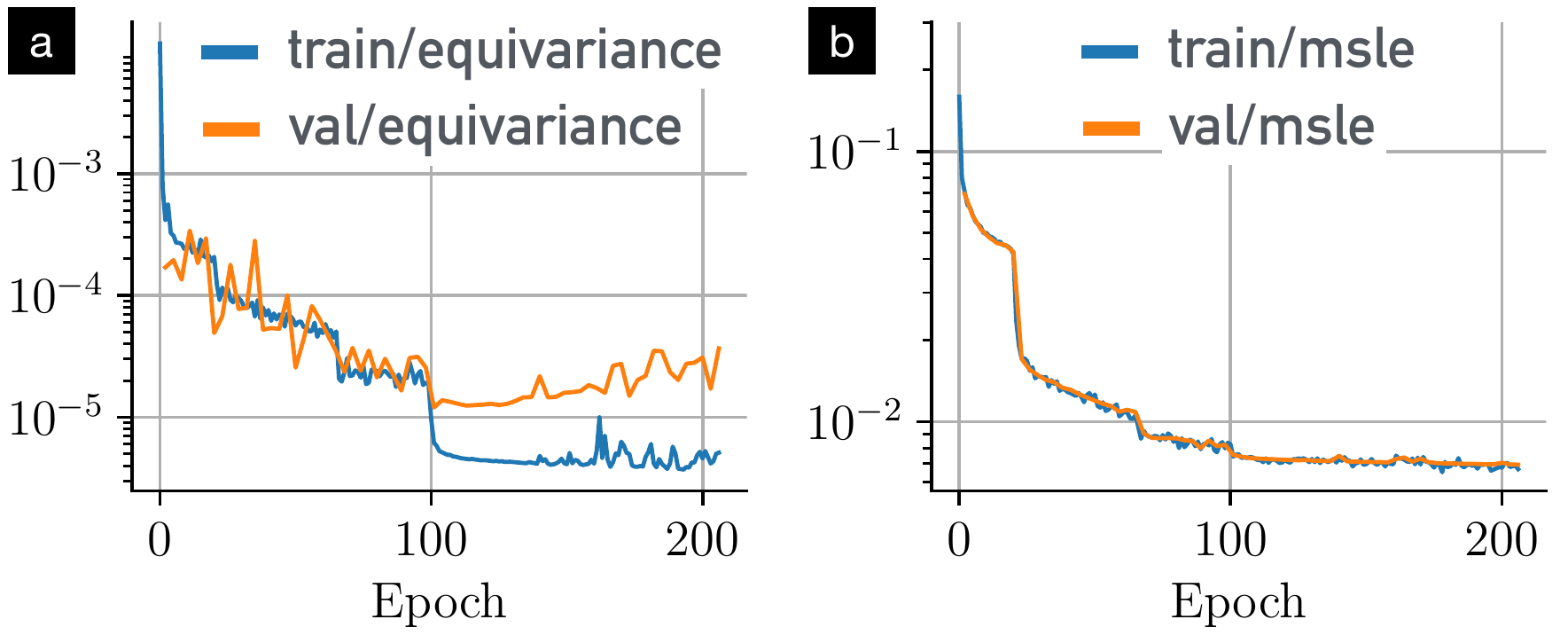}
    \caption{Loss curves during the training of the 2D model.}
    \label{fig:2d_training}
\end{figure}

The final loss is a weighted combination:
\begin{equation*}
    \mathcal{L} = \rho_{E}\mathcal{L}_{E} + \rho_{MSLE} \mathcal{L}_{MSLE},
\end{equation*}

where \(\rho_{E}/\rho_{MSLE}=25\). Ultimately for inference, we selected the model with the smallest MSLE validation error loss for our studies. In Fig.\ref{fig:2d_training}, training curves for the equivariance and MSLE losses for the training of the 2D model are shown. While the equivariance loss tends to overfit at some point, we have not encountered overfitting in the MSLE loss.

\subsection{Inversion}

For computing a potential from a density, we use gradient descent on the input through the model. The inversion for Fig.\ref{fig:bose_hubbard_2d}g uses 250 steps with a learning rate of 0.1.

The inverted potential \( \mu/U \) is restricted to the range \( (-1.5, 5.0)\) with a rescaled sigmoid function.

For the implementation of model architectures, training and inversion, we use Pytorch \cite{PyTorch}.

\end{document}